\documentstyle[12pt,aasms4]{article}

\received{December 18, 1996}
\accepted{January 31, 1997}
\slugcomment{to appear in ApJL} 

\lefthead{Chang \& Ho}
\righthead{RXTE Observation of PSR B1951+32}

\newcommand{\beqary}{\begin{eqnarray}}
\newcommand{\eeqary}{\end{eqnarray}}
\newcommand{\beq}{\begin{equation}}
\newcommand{\eeq}{\end{equation}}
\newcommand{\cm}{{\rm cm}}

\newcommand{\s}{{\rm s}}

\newcommand{\keV}{{\rm keV}}

\newcommand{\fu}{\cm^{-2}\s^{-1}\keV^{-1}}
\begin{document}

\title{RXTE Observation of PSR B1951+32}

\author{H.-K. Chang\altaffilmark{1} and C. Ho}
\affil{Los Alamos National Laboratory, MS D436, Los Alamos, NM 87545, USA}
\altaffiltext{1}{present address: Department of Physics, National
                 Tsing Hua University, Hsinchu 30043, Taiwan, ROC}

\begin{abstract}
We report results of RXTE observations of PSR B1951+32 using the
PCA instrument for 19k seconds during 1996 March 24th.
We applied the contemporaneous radio ephemeris and various
statistical tests to search for evidence of pulsation. These analyses
yield intriguing yet inconclusive evidence for the presence of the
pulsation in the time series: confidence level for the presence of
pulsation is 94.5\% in the 2.0-4.8 keV band and 97.6\% in the 4.8-6.3
keV band based on the $H$-test. Under the premise of non-detection of
pulsation, we derive estimated 2-$\sigma$ upper limits for the pulsed
flux to be $3.3\times 10^{-6}\,\fu$ in the 2.0-4.8 keV band,
$2.7\times 10^{-6}\,\fu$ in the 4.8-8.5 keV band, and $2.0\times
10^{-6}\,\fu$ in the 8.5-13.0 keV band.  These upper limits are
consistent with the trend of spectral turn-over from high-energy
gamma-ray emission as suggested by the OSSE upper limit.  Such
turn-over strongly suggests the outer magnetosphere as the emission
site for pulsed gamma-rays. These RXTE upper limits for X-ray
pulsation are, on the other hand, not consistent with the
extrapolation of reported power-law spectra from the point source
observed by ROSAT in the 0.1-2.4 keV band, assuming a constant pulse
fraction: The pulsed soft X-ray emission detected by ROSAT must follow
a much softer spectrum than that of the overall point source.

\end{abstract}


%*************************************************************************
\section{Introduction}
\label{intro}

The pulsar PSR B1951+32, located at the center of the morphologically
peculiar radio nebula CTB 80, is a 39.5-msec radio pulsar (Clifton et
al. 1987; Kulkarni et al. 1988) with a characteristic age of $1.1
\times 10^5$ yr and an inferred surface dipole magnetic field of $4.9
\times 10^{11}$ G.  An X-ray point source was observed within the
X-ray nebula related to the radio core of CTB 80 (Becker, Helfand \&
Szymkowiak 1982; Seward 1983; Wang \& Seward 1984).  Search for X-ray
pulsation from this source with EXOSAT yielded inconclusive evidence
for pulsation (confidence level of 97\%, by \"Ogelman \& Buccheri
1987, and 93\% by Angelini et al. 1988).  The pulsed emission was
detected by ROSAT at a 99\% confidence level (Safi-Harb, \"Ogelman \&
Finley 1995), which shows a single peak roughly consistent in phase
with the radio emission.  The overall spectrum of this point source in
0.1-2.4 keV is best fitted by a power law with a photon spectral index
$-1.6\pm 0.2$, and an estimated pulsed fraction of 0.35.

The EGRET instrument on CGRO observed and detected gamma-ray pulsation
from PSR B1951+32 above 100 MeV (Ramanamurthy et al. 1995), making it
a member of EGRET hard gamma-ray pulsar family with a similar age to
Vela and Geminga.  The gamma-ray lightcurve shows
two peaks at phase 0.16 and 0.60 with phase 0.0 being the radio peak.
Its spectrum in the EGRET energy range follows a power law with a photon
spectral index of about $-1.8$
(Ramanamurthy et al.\ 1995; Fierro 1995) over about
two decades of photon energy. Recently, pulsed emission is reported
from the COMPTEL instrument in the 0.75-10.0 MeV band (Kuiper et
al.\ 1996). The OSSE and BATSE instruments on CGRO only reported upper
limits of pulsed emission in the lower energy band (Schroeder et al.\
1995; Wilson et al.\ 1992).

There have been a number of models proposed to explain the gamma-ray
emission with dramatically different emission sites, some at or very
near the surface of the neutron star and some very far away.
Recently, Ho \& Chang (1996)
proposed a geometry-independent argument to constrain the possible
site and mechanism of the gamma-ray emission based on the commonality
of power-law emission in the EGRET (and possibly COMPTEL) pulsars. In
such arguments, it is important to know whether and how 
the gamma-ray power-law spectra turn
over towards low energy. (See Section 4 for
more discussions.)

To gain better understanding of the overall spectral behavior,
especially between keV and MeV, we conducted an observation of PSR
B1951+32 using both PCA and HEXTE on board RXTE during Cycle 1.
Analysis of the 19k-second PCA data does not yield conclusive evidence
for pulsation from 2.0 to 13.0 keV. The derived 2-$\sigma$ upper
limits provide support for the hard turn-over for the high-energy
gamma-ray emission. It also indicates that the soft X-ray pulsation
observed by ROSAT has a very soft spectrum.  We described the
observation in Section 2. The analyses and results for the PCA data
are discussed in Section 3. We discuss the theoretical implications of
this observation and future work in Section 4.

%*************************************************************************
\section{Observation}
\label{ob}  

The PCA and HEXTE on board RXTE were pointed at PSR B1951+32 on March
24, 1996 (MJD 50166), for about 10.5 hours including earth
occultations.  The RXTE mission, spacecraft and instrument
capabilities are described in Swank et al.\ (1995), Giles et al.\
(1995) and Zhang et al.\ (1993) The PCA consists of five essentially
identical PCUs with a total effective area of 6729 cm$^2$, with no
imaging capability. The field of view is one degree.

After examining the data, two exclusions were applied to the data
set. First, data from the PCA pulse-height channel 36-255 (13.0-90.0
keV) are excluded due to high instrumental noise. Second, we observed
unexplained anomalous increase during two intervals of our
exposure. Under the advice of RXTE Guest Observer Facility experts,
data obtained during these two intervals were excluded.  In the second half of
the observation, two of the five PCUs were turned off. The overall
usable data used for this analysis contain two segments of
$8928\,\s\,\times 6729\,\cm^2$ and $10304\,\s\,\times 4180\,\cm^2$ for
a total of $1.03\times 10^8\,\s\,\cm^2$, or equivalently, a total
integration time of 19232 seconds and an average effective area of
5363.3 cm$^2$.

Around the same epoch of the RXTE observation, PSR B1951+32 was also
monitored at Jodrell Bank Radio Observatory. The radio ephemeris is
summarized in Table 1 and used as the input for pulsation search.
%FFFFFFFFFFFFFFFFFFFFFFFFFFFFF
\placetable{ephemeris}
%FFFFFFFFFFFFFFFFFFFFFFFFFFFFF

%*************************************************************************
\section{Analysis and Results}
\label{ana}  

The data were reduced to the barycenter and analyzed using the JPL
DE200 ephemeris, the pulsar position listed in Table~\ref{ephemeris},
and standard RXTE reduction package (FTOOLS v.3.5.2 and XANADU/XRONOS
v.4.02). 
% Following standard procedure, instrumental and cosmic X-ray
%background subtraction and lightcurve extraction are performed with
%msec time resolution from four typical energy bands. 
% The average instrumental
%background count rate is roughly 1.5 counts/sec/PCU in band 1 and 2.5
%counts/sec/PCU in bands 2-4.
%  It is known that PCA channels 36-255
%have excessive instrumental noise (James Lochner, private
%communication). We have examined data from these channels and found
%the instrumental background to be 25 counts/sec/PCU. As a result, data
%from these channels are excluded from the analysis. 
%The cosmic X-ray
%background, estimated by the standard analysis is 3.5, 2.1, 3.0, and
%2.3 counts/sec for a total of five PCUs in bands 1, 2, 3, and 4
%respectively.
% The lightcurves are folded at the radio period and
%the appropriate time derivative to search for evidence of pulsation.
Lightcurve folding was performed separately for each of four
typical energy bands and various combinations using the radio ephemeris
in Table~\ref{ephemeris}.
The four typical energy
bands are commonly designated as band 1 through 4, with each covering
PCA channels 0-13 (2.0-4.8 keV), 14-17 (4.8-6.3 keV), 18-23 (6.3-8.5
keV), and 24-35 (8.5-13.0 keV), respectively. 
None of the folded lightcurves
showed significant deviation from a model steady distribution under
the Pearson's $\chi^2$-test (Leahy et al.\ 1983a,b). Specifically,
the $\chi^2$ values for the folded lightcurves shown in
Figure~\ref{lightcurve} are, for 19 degrees of freedom, 27.4 for band
1, 21.1 for band 2, and 8.38 for the combined band of 3 and 4. 
In addition to instrumental and cosmic X-ray background, the DC
component is mostly likely the contribution from the ROSAT point
source and its associated X-ray nebula.
%FFFFFFFFFFFFFFFFFFFFFFFFFFFFF
\placefigure{lightcurve}
%FFFFFFFFFFFFFFFFFFFFFFFFFFFFF

To further ascertain the absence of pulsation, we performed the
bin-independent parameter-free $H$-test (De Jager, Swanepoel \&
Raubenheimer 1989). In this analysis, all detected raw photons with
the corrected arrival time are used.
% The $H$-statistic is defined as
%\beq
%H\equiv\stackrel{\mbox{maximum}}{_{1\leq m\leq 20}}
%H\equiv{\max}_{(1\leq m\leq 20)}\,
%(Z_m^2 -4m+4)\, ,
%\eeq
%where $Z_m^2$ is the statistic from the $Z_m^2$-test (Buccheri et al.\
%1983), given by
%\beq
%Z_m^2\equiv\frac{2}{N}\sum_{k=1}^{m}\left[
%\left(\sum_{i=1}^{N}\cos 2\pi k\phi_{i}\right)^2 +
%\left(\sum_{i=1}^{N}\sin 2\pi k\phi_{i}\right)^2 
%\right]\, ,
%\eeq
%where $N$ is the total number of detected photons, and $\phi_i$ is the
%residual phase of the $i$th photon. The residual phase $\phi_i$
%relative to the known periodicity and its derivatives is
%\beq
%\phi_i=\mbox{fractional part of}\,(\nu(t_i-t_0)+\dot{\nu}(t_i-t_0)^2/2
%+\ddot{\nu}(t_i-t_0)^3/6)\, ,
%\eeq
%with $\nu$ being the pulsar frequency, $t_i$ the arrival time of the
%$i$th photon, and $t_0$ the radio epoch; $t_i$ and $t_0$ are measured
%at the solar system barycenter.  The probability distribution of the
%null hypothesis that the time series is generated by a steady source,
%is an exponential function of the $H$-statistic (De Jager, Swanepoel
%\& Raubenheimer 1989).
The $H$-test was applied to the data in different energy bands and
various combinations.  The results of the $H$-test all show high
probability that the data are consistent with steady source except for
band 1 and 2. The $H$-values are 7.286 and 9.334 for bands 1 and 2
(both at $m$=1). This yields a 5.5\% and 2.4\% probability of being
consistent with a steady source. Applying the straight $Z_1^2$-test
(the Rayleigh test, which is more appropriate if the underlying pulse
profile is sinusoidal), the probability of the data being consistent
with a steady source is 2.9\% and 0.9\% for bands 1 and 2.

Based on these analyses, we do not consider the null probability, although
intriguingly small using the $H$-test, provides conclusive evidence of
pulsation from PSR B1951+32 in the XTE/PCA energy band.

The upper limit of pulsed flux is estimated following the prescription
given by Ulmer et al. (1991).
%\beq
%\sigma=\frac{f^{1/2}C_t^{1/2}}{AT\triangle E}\, ,
%\eeq
%where
%\beq
%f=\frac{\beta}{1-\beta}\, ,
%\eeq
%and $\beta$ is the duty cycle of the pulse, $C_t$ the total number
%of counts, $AT$ the total exposure, and $\triangle E$ the
%$width of the energy band.
Assuming a duty cycle of 0.5 and combining bands 2 and 3 to yield a
comparable total number of counts to those in bands 1 and 4 individually, we
obtain the following 2-$\sigma$ upper limits, which are also shown
in Figure~\ref{spectrum}: $3.3\times 10^{-6}\,\fu$ for 2.0-4.8 keV,
$2.7\times 10^{-6}\,\fu$ for 4.8-8.5 keV, and $2.0\times 10^{-6}\,\fu$
for 8.5-13.0 keV.
%FFFFFFFFFFFFFFFFFFFFFFFFFFFFF
\placefigure{spectrum}
%FFFFFFFFFFFFFFFFFFFFFFFFFFFFF

%*************************************************************************
\section{Discussion}
\label{dis}  

{\it Gamma-ray pulsar:} The current XTE/PCA upper limits provide
support to the combined CGRO observation that the gamma-ray pulsed
emission from PSR B1951+32 follows a power law with a significant
break towards low energy (EGRET and COMPTEL detection along with OSSE
and BATSE upper limits), as indicated in Figure~\ref{spectrum}. Such
spectral behavior is seen in the Vela and Geminga pulsars. For PSR
B1951+32, the break energy (photon energy at which there is a
significant break in photon spectral indices to, say, harder than $-1$) is
estimated to be between 70 keV and 3 MeV. As noted in Ho \& Chang
(1996), this common trait could play an important role in the
theoretical modeling of this family of ``EGRET pulsars.''  The power
law of these pulsars typically covers two orders of magnitude in the EGRET
band with best-fit photon spectral indices in the range of $-1.4$ to
$-1.8$. These photon spectral indices cannot be produced by a mono-energetic
relativistic electron distribution under currently proposed radiation
mechanisms. The most likely origin of this power-law behavior is the
cooling (energy loss) through the dominant radiation mechanism
responsible for the gamma-rays. For example, the most simplistic
cooling model for a steady state electron distribution will yield a
photon spectral index of $-3/2$ for synchrotron radiation and $-5/3$ for
curvature radiation. The cooling spectrum will continue towards low
energy until the electron distrbution (in energy space) is no longer
affected by cooling: i.e. the cooling spectrum turns hard at the break
energy which corresponds to the location and electron energy where the
radiative cooling time scale is comparable to the dynamical time scale
of the relativistic motion of the electrons. Such an argument allows us
to constrain the radiation mechanism and, more importantly, the
emission site, which to date remains unsettled with great bifurcation
among gamma-ray pulsar models. Following this argument and examining
various radiation mechanisms, we find that, for PSR B1951+32, the
gamma-ray pulsations are most likely generated by synchrotron
radiation with a typical pitch angle of 0.1 to 0.001 and emission site
is about $10^7$ to $10^8$ cm from the star: i.e. in the outer
magnetosphere.

{\it X-ray pulsar:} Safi-Harb, \"Ogelman \& Finley (1995) reported
ROSAT observation of PSR B1951+32 with an estimated pulsed fraction of
0.35 and an overall spectrum following a power law with a photon spectral
index of $-1.6\pm 0.2$. To date, there is no published spectrum for the
pulsed component. Safi-Harb et al. (1995) estimated the soft X-ray
pulsation duty cycle to be 0.1. Such a duty cycle will lower the XTE
upper limit estimated above. It is clear from Figure~\ref{spectrum}
that the XTE/PCA observation necessitates a steep drop-off at around 2 keV
for the pulsed component. This is consistent with Safi-Harb et al.'s
report of no pulsation near 2 keV.  It is almost certain that, for PSR
B1951+32, the ROSAT observed pulsation below 1.3 keV is separate from
the EGRET/COMPTEL observed gamma-ray pulsation. More than likely, they
are from different origin and emitted at different location.
% The sharp
%drop-off at 2 keV is consistent with the common assumption that these
%pulsed soft X-rays come from hot polar caps. For the purpose of
%discussion, we take the polar cap to have a temperature corresponding
%to $3kT=1.3$ keV, with 1.3 keV being the energy band with the greatest
%pulsation significance.  Application of the XTE/PCA upper limit
%requires the projected area of the hot spot to be less than $2.5\times
%10^9\,(d/{2.5}{\rm kpc})\!^2\,\cm^2$, where $d$ is the distance to the
%pulsar.  This upper limit of the projected area is within the limit of
%the theoretical polar cap size for PSR B1951+32.  However, the flux
%from such a hot spot cannot account for the reported 0.35 pulsed
%fraction.  This issue should be resolved with a better
%characterization of the pulsed spectra in the ROSAT band.

In summary, XTE/PCA observation over 19k seconds shows no definitive
evidence of
pulsation. The upper limits can be used to help constrain theoretical
models. Our analysis does show tantalizing hints. In addition to the
small null probability from the $H$-test, the peaks for bands 1 and 2
in Figure~\ref{lightcurve}, taken at face value, are separated by 0.45
in phase, reminiscent to that of the Geminga pulsar in 0.07-1.5 keV
(Halpern \& Ruderman 1993).  A long exposure on XTE, e.g. 100 ksecs,
will provide better statistics and help advance our understanding of
this and other similar pulsars.

%*************************************************************************
%\section{Summary}
%\label{sum}  

%-------------------------------------------------------------------------

%*************************************************************************
\acknowledgments
We thank the XTE team in XTE/GOF at GSFC, especially James Lochner, for
their help in data reduction and analysis. We also thank Andrew Lyne for
providing
the radio ephemeris and comments on the manuscript.  
Many useful discussions
with Ed Fenimore and James Theiler are gratefully acknowledged.
We are appreciative of the referee, F. Seward, for his many helpful
comments in improving this paper.
This
work was performed under the auspices of the US Department of Energy
and was supported in part by the RXTE Guest Observer Program and CGRO
Guest Investigator Program.

%*************************************************************************

%**************************
\clearpage

\figcaption[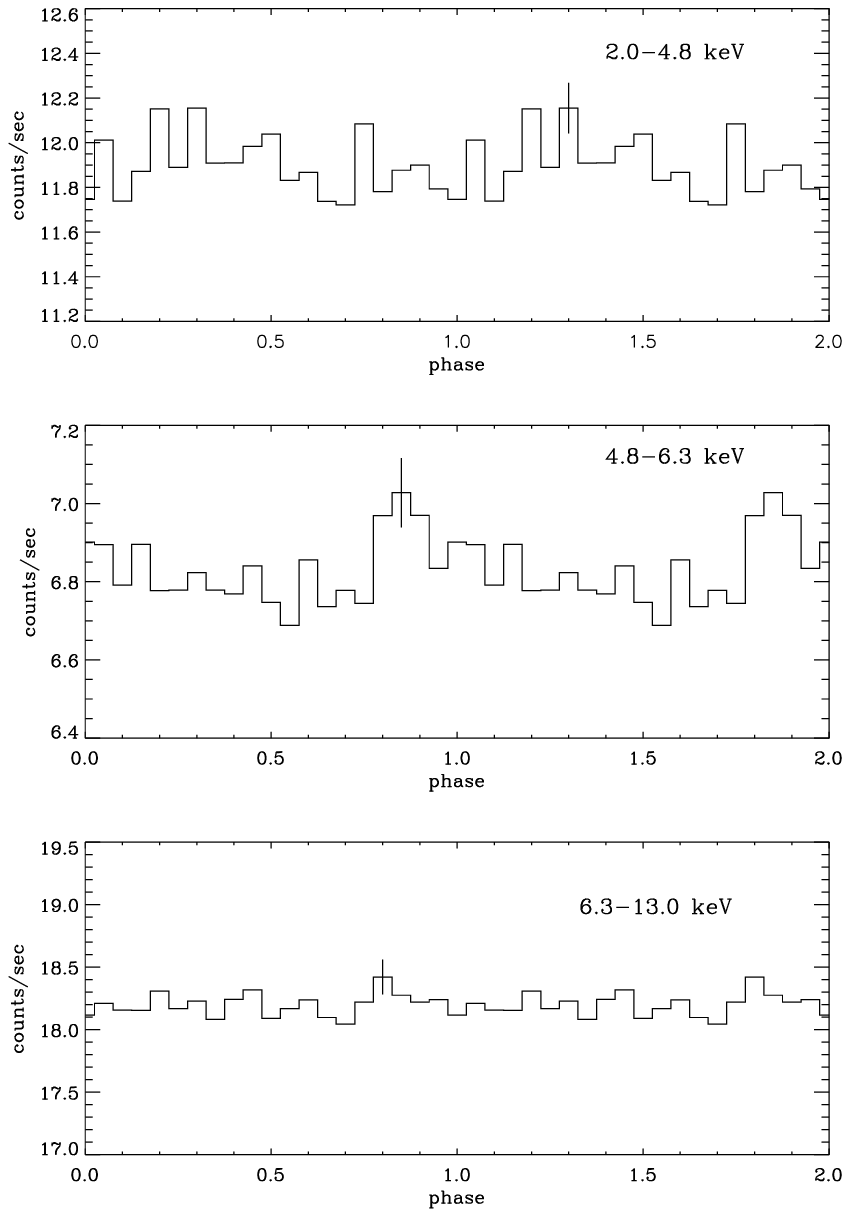]{
Epoch-folded lightcurves of PSR B1951+32 in three
different energy bands. Two periods are shown for clarity.
Phase 0 is referred to the radio peak.
In addition to instrumental and cosmic X-ray background, the DC
component is mostly likely the contribution from the ROSAT point
source and its associated X-ray nebula.
\label{lightcurve}
}

\figcaption[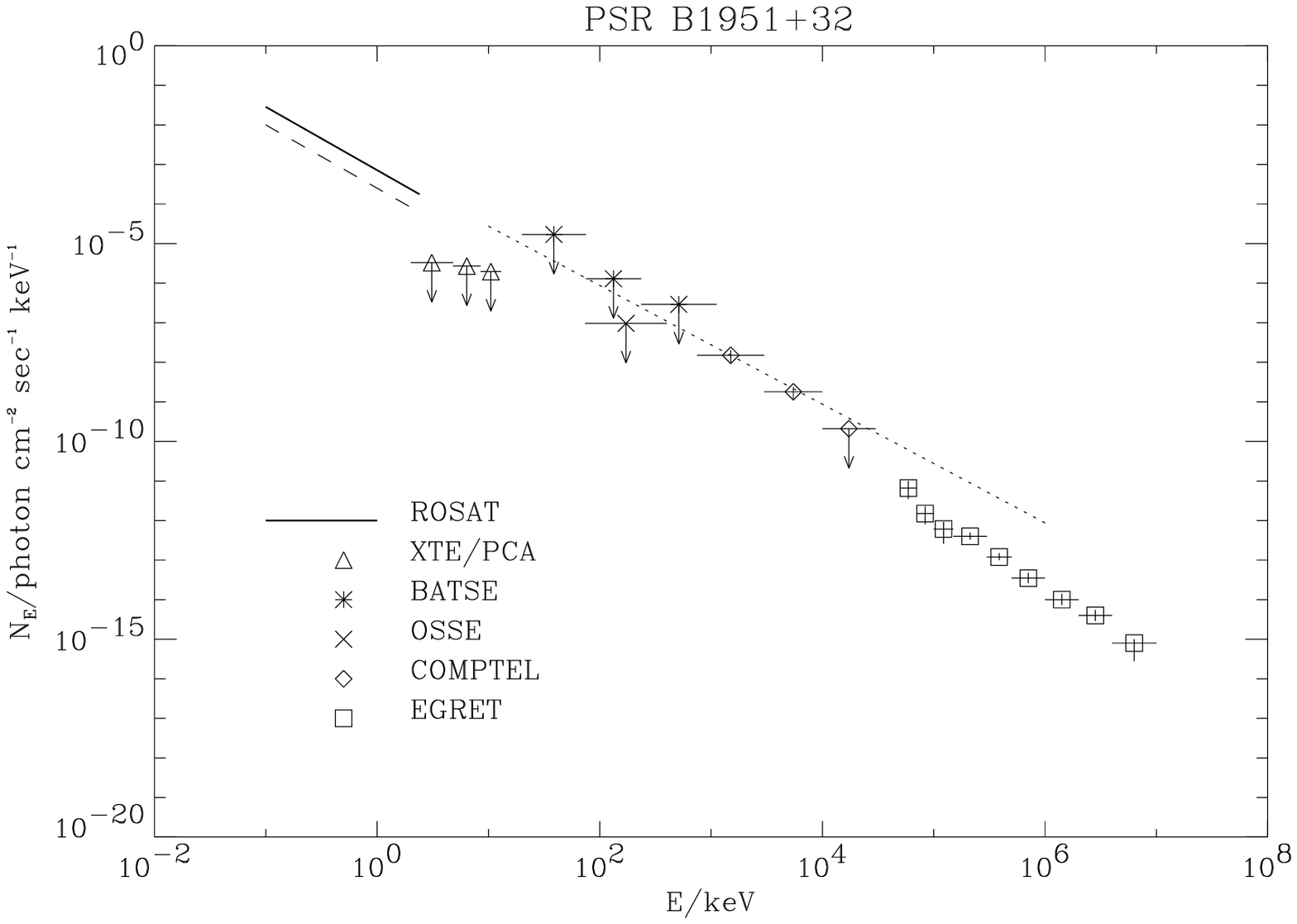]{
The spectrum of PSR B1951+32 from 0.1 keV up to 10 GeV.
The ROSAT data is for the X-ray point source including pulsed and
unpulsed components. The dashed line is the solid line times
0.35, the reported pulsed fraction of this point source.
The dotted line is a power law with a photon spectral index $-3/2$
passing through
the COMPTEL data point in 0.75-3.0 MeV.
Upper limits are at a 2-$\sigma$ level for XTE/PCA, OSSE, and COMPTEL,
and 3-$\sigma$ for BASTE.
References: EGRET -- Fierro (1995); COMPTEL -- Kuiper at al.\ (1996);
OSSE -- Schroeder et al.\ (1995); BATSE -- Wilson et al.\ (1992);
ROSAT -- Safi-Harb, \"Ogelman \& Finley (1995).
\label{spectrum}
}

\clearpage

\begin{deluxetable}{llr}
%\footnotesize
\tablecaption{
Radio ephemeris of PSR B1951+32\tablenotemark{a}
\label{ephemeris}}
%\tablewidth{0pt}
\tablehead{
\colhead{} & \colhead{} & \colhead{} 
} 
\startdata
Validity interval & (MJD) & 50057 - 50207 \nl
Epoch, $t_0$ & (MJD) & 50132.000000201 \nl
$\alpha_{2000}$ & & $19^{\rm h}52^{\rm m}58^{\s}.276$ \nl
$\delta_{2000}$ & & $32^{\circ}52'40''.68$ \nl
$\nu$ & (Hz) & 25.2963865292632 \nl
$\dot{\nu}$ & (Hz/s) & $-3.74022\times 10^{-12}$ \nl
$\ddot{\nu}$ & (Hz/s/s) & $2.81\times 10^{-22}$ \nl 
\enddata
\tablenotetext{a}{provided by Andrew Lyne (1996, private communication)}
\end{deluxetable}

\clearpage
\plotone{lightcurve.eps}
\clearpage
\plotone{spectrum.eps}
%**************************
\end{document}